\pdfoutput=1
\documentclass [sigconf,natbib=true]{acmart}

\AtBeginDocument{%
  }

\copyrightyear{2023}
\acmYear{2023}
\setcopyright{acmlicensed}\acmConference[SIGIR '23]{Proceedings of the 46th
International ACM SIGIR Conference on Research and Development in
Information Retrieval}{July 23--27, 2023}{Taipei, Taiwan}
\acmBooktitle{Proceedings of the 46th International ACM SIGIR Conference on
Research and Development in Information Retrieval (SIGIR '23), July 23--27,
2023, Taipei, Taiwan}
\acmPrice{15.00}
\acmDOI{10.1145/3539618.3592070}
\acmISBN{978-1-4503-9408-6/23/07}

\usepackage[utf8]{inputenc} 
\usepackage[T1]{fontenc}    
\usepackage{url}            
\usepackage{nicefrac}       
\usepackage{microtype}      
\usepackage{xcolor}         
\usepackage{bm}
\usepackage{bbm}
\usepackage{todonotes}
\usepackage{algorithm}
\usepackage{algpseudocode}
\usepackage{multicol}
\usepackage{caption}
\usepackage{multirow}
\usepackage{makecell}
\usepackage{booktabs}           
\usepackage{amsfonts}           
\usepackage{graphicx}           
\usepackage{duckuments}         
\usepackage{bold-extra}
\usepackage{amsmath}

\DeclareMathOperator*{\argmin}{arg\,min}

\begin{document}

\title{The Dark Side of Explanations: Poisoning Recommender Systems with Counterfactual Examples} 

\author{Ziheng Chen}
\email{ziheng.chen@stonybrook.edu}
\orcid{0000-0002-2585-637X}
\affiliation{%
  \institution{Stony Brook University, NY, USA}
  \country{}
}

\author{Fabrizio Silvestri}
\email{fsilvestri@diag.uniroma1.it}
\orcid{0000-0001-7669-9055}
\affiliation{
  \institution{Sapienza University of Rome, Italy}
  \country{}
}

\author{Jia Wang}
\email{jia.wang02@xjtlu.edu.cn}
\orcid{0000-0002-3165-7051}
\affiliation{%
  \institution{The Xi'an Jiaotong-Liverpool
University, Suzhou, China}
  \country{}
}

\author{Yongfeng Zhang}
\email{yongfeng.zhang@rutgers.edu}
\orcid{0000-0002-1243-1145}
\affiliation{%
  \institution{Rutgers University, NJ, USA}
  \country{}
}

\author{Gabriele Tolomei}
\email{tolomei@di.uniroma1.it}
\orcid{0000-0001-7471-6659}
\affiliation{
  \institution{Sapienza University of Rome, Italy}
  \country{}
}

\renewcommand{\shortauthors}{Chen et al.}

\begin{abstract}
Deep learning-based recommender systems have become an integral part of several online platforms. However, their black-box nature emphasizes the need for explainable artificial intelligence (XAI) approaches to provide human-understandable reasons why a specific item gets recommended to a given user.
One such method is \textit{counterfactual explanation} (CF).
While CFs can be highly beneficial for users and system designers, malicious actors may also exploit these explanations to undermine the system's security.
\\
In this work, we propose H-CARS, a novel strategy to poison recommender systems via CFs. Specifically, we first train a logical-reasoning-based surrogate model on training data derived from counterfactual explanations. By reversing the learning process of the recommendation model, we thus develop a proficient greedy algorithm to generate fabricated user profiles and their associated interaction records for the aforementioned surrogate model.
Our experiments, which employ a well-known CF generation method and are conducted on two distinct datasets, show that H-CARS yields significant and successful attack performance.
\end{abstract}




\maketitle

\section{Introduction}
\label{sec:intro}

The past few decades have witnessed the tremendous success of deep learning (DL) techniques in personalized recommender systems.
Indeed, DL-based recommender systems overcome the obstacles of conventional models and achieve state-of-the-art performance \cite{li2021hausdorff,wang2020cdlfm}. 
Despite that, they suffer from a lack of transparency and explainability. This issue may limit their deployment, especially in some critical domains, considering the increasing demand for explainable artificial intelligence (XAI) worldwide~\cite{tolomei2017kdd,tolomei2021tkde,mothilal2020fat,karimi2020aistats,le2020grace,lucic2019focusAAAI,siciliano2022newron,chen2022cikm}.

Recently, \textit{counterfactual explanation} (CF) has emerged as a key tool to attach motivation behind items recommended to users.
Concretely, CFs generate a minimal set of meaningful interactions, without which the recommended items will not end up in the list of suggestions for specific users.
While much prior research shows how counterfactual explanations can enhance recommender systems' transparency, few studies investigate their potential security and privacy threats. 
Instead, several studies have shown possible hazards of CFs for the classification task~\cite{aivodji2020model,li2021tackling,pawelczyk2022privacy,zhao2021exploiting,duddu2022inferring}.
For example, Aïvodji et al.~\cite{aivodji2020model} demonstrate that a high-fidelity model could be extracted by detecting the information of decision boundaries embedded in CFs. This finding inspired DualCF~\cite{wang2022dualcf}, which leverages CFs and their counterfactual explanations to overcome decision boundary shift issues. Additionally, Pawelczyk et al.~\cite{pawelczyk2022privacy} conduct the membership inference attacks by using the distance between the data instances and their corresponding CFs. 

This work investigates possible risks induced by CFs for the recommendation task.
Specifically, we demonstrate how an adversary can use CFs to conduct poisoning attacks on black-box recommender systems. 
Technically, we first train a logical reasoning model as a surrogate by exploiting CFs. Then, a limited number of controlled users with fake crafted interactions are designed by matching the optimal representation of a target recommended item via an optimization framework. 
We refer to our attack strategy as H-CARS (Horn-Clause Attacks to Recommender Systems).

Overall, the main contributions of our work are as follows:
\begin{itemize}
\item 
We unveil security issues of applying counterfactual explanations in recommender systems and provide the first study of poisoning attacks for recommendation via CFs.
\item 
We jointly model logical and counterfactual reasoning by leveraging CFs and partial factual interactions to enhance the surrogate model's performance. A counterfactual loss $\mathcal{L}_{cf}$ is proposed to highlight the necessary items while mitigating spurious correlations.
\item 
We propose a novel poisoning attack for neural logical-based recommender systems. Specifically, inspired by~\cite{zhang2021data}, we reverse the traditional attacking optimization procedure where, instead of re-training the recommendation model, we start from computing the optimal item embeddings, which we leverage to find the fake user-item interactions.
\item 
We conduct experiments on two real datasets to analyze the attacking performance.
\end{itemize}

The remainder of this paper is organized as follows. 
We provide background and preliminaries in Section~\ref{sec:background}. 
In Section~\ref{sec:problem}, we present the attack model and describe our proposed method H-CARS, in Section~\ref{sec:method}. We validate H-CARS in Section~\ref{sec:experiments}.
Finally, we conclude our work in Section~\ref{sec:conclusion}.

\section{Background and Preliminaries}
\label{sec:background}
We consider the standard collaborative filtering recommendation task.
Let $\mathcal{U}=\{u_1,\ldots,u_m\}$ be a set of $m$ users, and $\mathcal{I}=\{i_1,\ldots,i_n\}$ be a set of $n$ items.
We represent the \textit{factual} interactions between users and items with a binary user-item matrix $\mathcal{Y}\in\{0,1\}^{m\times n}$, where $Y_{u,i}=y_{u,i} = 1$ indicates that user $u$ has interacted with item $i$, or $0$ otherwise. Specifically, we denote the interaction history of a user $u$ as $\mathcal{I}_{u}=\{i\in \mathcal{I}~|~y_{u,i} = 1\}$. 

We assume there exists a recommendation model $f$ that can estimate the value $\hat{y}_{u,i}$ for each $u\in \mathcal{U}$ and each $i\in \mathcal{I}$, such that $i\notin \mathcal{I}_u$, as $\hat{y}_{u,i} = f(\bm{h}_u, \bm{h}_i)$, where $\bm{h}_u,\bm{h}_i\in \mathbb{R}^d$ are suitable user and item representations, respectively, and $f: \mathbb{R}^d \times \mathbb{R}^d \mapsto [0,1]$ is a function that measures the preference
score for user $u$ on item $i$.\footnote{A similar reasoning would apply if we instead considered explicit ratings, i.e., $f: \mathbb{R}^d \times \mathbb{R}^d \mapsto \mathbb{R}$.}
The score computed with $f$ is used to rank items to populate a list $\hat{\mathcal{I}}^k_{u}\subseteq \mathcal{I}\setminus \mathcal{I}_u$ of the top-$k$ recommended items for user $u$, i.e., the list of $k$ unrated items most likely relevant to user $u$ according to $f$.

Thus, given a target item $t$ recommended to a user $u$ by $f$ (i.e., $t\in \hat{\mathcal{I}}^k_{u}$), its counterfactual explanations are the minimal subset of $u$'s historical item interactions $\mathcal{I}_u$ whose removal results in evicting $t$ from the top-$k$ list of recommendations $\hat{\mathcal{I}}^k_{u}$, namely $\mathcal{I}_{u,t}^{CF} \subset \mathcal{I}_u$.

\section{Attack Model}
\label{sec:problem}

Our strategy involves poisoning attacks that aim to promote target items to legitimate users by manipulating interactions of controlled users. Formally, let $\mathcal{T}\subseteq \mathcal{I}$ denote the set of target items, $\mathcal{U}$ and $\mathcal{U}'$ denote the set of legitimate users and users controlled by the attacker, respectively. 
Accordingly, $\mathcal{Y}$ and $\mathcal{Y}'$ denote the factual and fake interactions. 
In practice, online recommender systems limit access to the full user-item interaction data. Our attack model assumes that the factual interaction matrix $\mathcal{Y}$ can only be partially observed. Recommender systems are also typically considered "black-boxes," with the model's internals being inaccessible to attackers. However, recommender systems often have a public API that can be queried to generate a top-$k$ list of suggestions for a user $u$, which an attacker can use. Additionally, an explanation API for the recommendation model can be exploited to produce CFs for any user and target item pair $(u,t)$.

To quantify the power of the attack~\cite{han2022solution}, we define its success rate $HR(t)$ -- the hit ratio of item $t\in \mathcal{T}$ -- as the probability that $t$ appears in the top-$k$ list of recommended items for an actual user, i.e., $\mathbb{P}(\{t\in \hat{\mathcal{I}}_u^k~|~u\in \mathcal{U}\})$. 

The attacker aims to create fake interactions $\mathcal{Y}'$ for $\mathcal{U}'$, with the objective of maximizing the hit ratio of the target items. To formalize this problem, we follow prior work~\cite{zhang2021data}\cite{huang2021data}\cite{li2020dynamic} and present a bi-level optimization formulation:
\begin{equation}
\label{eq:Globalattack}
\begin{aligned}
 \max &
 \sum_{t \in \mathcal{T}}HR(t)\\
\textrm{subject to: } & \bm{\theta}^{*}=\argmin\limits_{\bm{\theta}}\Big\{\mathcal{L}_{train}(\mathcal{Y},\hat{\mathcal{Y}}_{\bm{\theta}})+\mathcal{L}_{train}(\mathcal{Y}',\hat{\mathcal{Y}}'_{\bm{\theta}})\Big\},
\end{aligned}
\end{equation}
where $\bm{\theta}$ denotes the model parameters, and $\mathcal{L}_{train}$ denotes the training loss of the target recommendation model $f$. $\mathcal{Y}_{\bm{\theta}}$ and $\mathcal{Y}'_{\bm{\theta}}$ are predictions from the model trained on $\mathcal{Y} \cup \mathcal{Y}'$.

\section{Proposed Method: H-CARS}
\label{sec:method}

\begin{figure}
    \centering
    \includegraphics[width=\columnwidth]{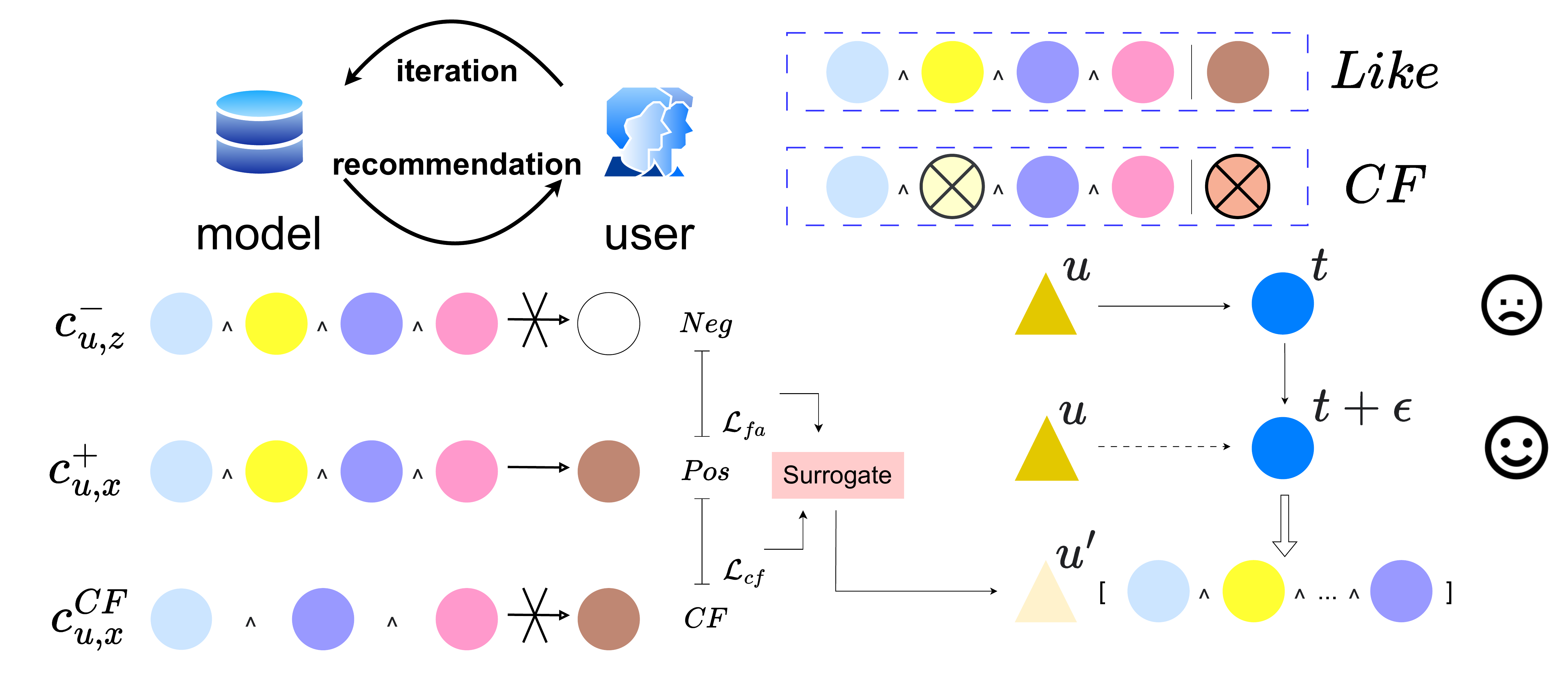}
    \caption{Our proposed H-CARS poisoning attack framework.}
 \vspace{-0.5cm}   \label{fig:h-cars}
\end{figure}

This section describes H-CARS (Horn-Clause Attacks to Recommender Systems), our poisoning method for recommender systems, via counterfactual explanations, depicted in Fig.~\ref{fig:h-cars}. The attacker generally cannot access the target model's internals, so it cannot optimize~(\ref{eq:Globalattack}) directly.
A common practice in black-box scenarios is first training a surrogate model $\hat{f}\approx f$ to generate fake user-item interactions that can be used to poison the target system $f$.

\subsection{Extracting Logical Reasoning Models with Counterfactual Explanations}

To build a surrogate model that effectively utilizes the API of CF reasoning, it is crucial to capture the inherent logic of the targeted black-box recommender system. To this end, we adopt the Neural Collaborative Reasoning (NCR) framework as the surrogate recommendation model in this study. NCR encapsulates logical reasoning into a dynamic neural architecture in Horn clause form (hence the name of our method). More specifically, NCR encodes a user-item interaction $(u,j)$ into an \textit{event vector}:
\begin{equation}
\boldsymbol{e}_{u,j}=\boldsymbol{W}_2\phi(\boldsymbol{W}_1(u,j)+b_1)+b_2,
\end{equation}
where $j\in \mathcal{I}_u$ and, with a slight abuse of notation, we refer to $u$ and $j$ as their corresponding embedding vectors ($\bm{h}_u$ and $\bm{h}_j$, respectively). 
Moreover, $\boldsymbol{W}_1$, $\boldsymbol{W}_2$, $b_1$ and $b_2$ are event parameters that need to be learned, whereas $\phi$ is the well-known ReLU activation function.

To transform the user-item interactions into a logical expression, basic neural logical modules, AND$(\land)$, OR$(\lor)$, NOT$(\lnot)$, are introduced. Ideally, predicting if an item $x$ is to be recommended to user $u$ based on the interaction history is equal to deciding whether the following Horn clause is \texttt{true} or \texttt{false}:
\begin{equation}
    \boldsymbol{e}_{u,1} \land \boldsymbol{e}_{u,2} \land \cdots \land \boldsymbol{e}_{u,j} \Rightarrow \boldsymbol{e}_{u,x},
\end{equation}
where $i\in \mathcal{I}_u~\forall i \in\{1,2,\ldots,j\}$.
Based on the definition of material implication, the above statement is equivalent to the following logic expression vector:
\begin{equation}
   \boldsymbol{c}_{u,x}^{+}=(\lnot \boldsymbol{e}_{u,1} \lor \lnot \boldsymbol{e}_{u,2} \lor \cdots \lor \lnot \boldsymbol{e}_{u,j})\lor \boldsymbol{e}_{u,x}.
\end{equation}
The recommendation score of $x$ is calculated based on the similarity between the logic expression vector $\boldsymbol{c}_{u,x}^{+}$ and the constant \texttt{true} vector denoted by $\bm{1}$.
We adopt the pair-wise learning algorithm for the factual interaction for model training. Specifically, we construct negative expression: 
\begin{equation}
   \boldsymbol{c}_{u,z}^{-}=(\lnot \boldsymbol{e}_{u,1} \lor \lnot \boldsymbol{e}_{u,2} \lor \cdots \lor \lnot \boldsymbol{e}_{u,j})\lor \boldsymbol{e}_{u,z},
\end{equation}
by sampling an item $z\in \mathcal{I}\setminus \mathcal{I}_u$ that $u$ did not interact with.
Then, we opt for the Bayesian Personalized Ranking (BPR) loss:
\begin{equation}
  \mathcal{L}_{fa}=-\sum_{u \in \mathcal{U}}\sum_{x \in \mathcal{I}_{u}^+}\sum_{z \in \mathcal{I}_{u}^-}\mbox{log}(sim(\boldsymbol{c}_{u,x}^{+},\bm{1})-sim(\boldsymbol{c}_{u,z}^{-},\bm{1})),
\end{equation}
where $\mathcal{I}_{u}^{+}$ and $\mathcal{I}_{u}^{-} = \mathcal{I}\setminus \mathcal{I}_u$ represents the positive and negative item sets for user $u$, respectively, and $sim$ measures the similarity between two vectors (e.g., cosine similarity). 

Since our partial factual interaction data is limited, CFs generated by the API can be used for data augmentation while alleviating spurious correlations~\cite{li2022mitigating}\cite{mu2022alleviating}\cite{kong2021learning}. For an item $x$, after subtracting $\mathcal{I}_{u,x}^{CF}$ from $\mathcal{I}_u$, the remaining set $\mathcal{I}_u\setminus \mathcal{I}_{u,x}^{CF}$ leads to a removal of item $x$ from the top-$k$ list of suggestions for user $u$, i.e., $x\notin \hat{\mathcal{I}}_u^k$ anymore. Logically, the counterfactual expression below tends to be \texttt{false}:
\begin{equation}
 \boldsymbol{c}_{u,x}^{CF}= \Bigg(\bigvee_{j\in \mathcal{I}_u\setminus \mathcal{I}_{u,x}^{CF}}\lnot \boldsymbol{e}_{u,j}\Bigg)\lor \boldsymbol{e}_{u,x}
\end{equation}

For the generated counterfactual interaction, its expression vector $\boldsymbol{c}_{u,x}^{CF}$ is expected to be very distinct from $\boldsymbol{c}_{u,x}^{+}$, to weaken the spurious correlations by contrasting the difference. Hence, the counterfactual loss could be written as:
 \begin{equation}
\mathcal{L}_{cf}=-\sum_{u \in \mathcal{U}{U}}\sum_{x \in \mathcal{I}_{u}^{+}}\mbox{log}(sim(\boldsymbol{c}_{u,x}^{+},\boldsymbol{c}_{u,x}^{CF})).
\end{equation}

We integrate all the losses together to achieve the final objective:
 \begin{equation}
\mathcal{L}=\mathcal{L}_{fa}+\lambda_1\mathcal{L}_{cf}+\lambda_2\mathcal{L}_{reg},
\end{equation}
where $\mathcal{L}_{reg}$ is a regularization term that ensures the logical module satisfies the logical laws~\cite{chen2021neural}\cite{jin2021towards}.

\subsection{Poisoning Logical Reasoning Models}
Instead of constructing $\mathcal{Y}'$ and $\mathcal{U}'$ directly by solving~(\ref{eq:Globalattack}), we start from the goal and backtrack the optimization process~\cite{zhao2016group}\cite{zhang2021data}. Consider a simple case where the attacker wants to promote the target item $t$ to a user $u$ with interaction history $\mathcal{I}_{u}$. To achieve this,  we want the result of the following logic chain to be \texttt{true}:
\begin{equation}
    \boldsymbol{c}_{u,t}^{+}=(\lnot \boldsymbol{e}_{u,1} \lor \lnot \boldsymbol{e}_{u,2} \lor \cdots \lor \lnot \boldsymbol{e}_{u,j})\lor \boldsymbol{e}_{u,t}.
\end{equation}

Since the attacker cannot modify observed interactions of legitimate users in the training set, we focus on leveraging the controlled users to manipulate the embedding of $t$. Formally, we first determine the optimal item embedding $t+\epsilon$ by maximizing:
\begin{equation}
\varepsilon_t^{*}=\mbox{argmax}_{\varepsilon} sim\Big((\lnot \boldsymbol{e}_{u,1} \lor \lnot \boldsymbol{e}_{u,2} \lor \cdots \lor \lnot \boldsymbol{e}_{u,j})\lor \boldsymbol{e}_{u,t+\varepsilon}, \bm{1}\Big)
\end{equation}
Then, we transform the problem of promoting target item $t$ into the problem of shifting embedding $t$ to $t+\varepsilon$. In other words, we need to ensure the sum of terms involving item $t$ in the training loss decrease after the shifting.

\begin{equation}
-\sum_{u \in mathcal{U} \cup \mathcal{U}'}{sim}(\boldsymbol{c}_{u,t+\varepsilon}^{+},\bm{1}) \leq -\sum_{u \in \mathcal{U} \cup \mathcal{U}'}{sim}(\boldsymbol{c}_{u,t}^{+},\bm{1}).
\end{equation}

Here, we utilize the training loss of the surrogate to emulate the optimization process of the original recommendation model. 
Since the attacker can only control users in $\mathcal{U}'$, their goal is to generate the optimal controlled user $m^{*}\in \mathcal{U}'$  and its interaction history $\mathcal{I}_{{m}*}$ that minimize the following loss function in each iteration,
\begin{equation}
\label{eq:fake-users}
    m^{*},\mathcal{I}_{{m}*}=\mbox{argmax}_{m,\mathcal{I}_{m}}{sim}(\boldsymbol{c}_{m,t+\varepsilon}^{+},\bm{1}).
\end{equation}
Practically, we iteratively generate the controlled users. In each iteration, we find the optimal direction of item perturbation by jointly considering the target items $t \in \mathcal{T}$. Then, according to~(\ref{eq:fake-users}), the new $m^{*}$ and $\mathcal{I}_{{m}*}$ are obtained and included into $\mathcal{U}'$ and  $\mathcal{Y}$.

\section{Experiments}
\label{sec:experiments}
\begin{table*}[ht!]
.\caption{\label{tab:Result80}
$HR@10\times100$ for different attacks with $80\%$ training data (left) and $30\%$ training data (right) on two datasets.
}

\scalebox{0.75}{
\centering
\begin{tabular}{lll|cccc|}
\cline{4-7}
& & & \multicolumn{4}{c|}{{\bf Percentage of Cotrolled Users}}
\\
\hline
 \multicolumn{1}{|l|}{\bf Dataset} & \multicolumn{1}{l}{\thead{Recommendation\\Model}} & \multicolumn{1}{|l|}{\thead{Attack\\Method}} & \multicolumn{1}{c|}{\thead{$0.5\%$}} & \multicolumn{1}{c|}{\thead{$1\%$}} & \multicolumn{1}{c|}{\thead{$3\%$}} & \multicolumn{1}{c|}{\thead{$5\%$}}
\\
\hline
\multicolumn{1}{|l|}{\multirow{8}{*}{\em MovieLens}} & 
\multicolumn{1}{l}{\multirow{4}{*}{NCF}} & 
\multicolumn{1}{|l|}{{DL-Attack}} & 
\multicolumn{1}{c|}{$0.34$} & 
\multicolumn{1}{c|}{$0.39$} &
 \multicolumn{1}{c|}{$0.72$} & 
\multicolumn{1}{c|}{$0.82$} 
\\
\cline{3-7}
\multicolumn{1}{|l|}{} &
&
\multicolumn{1}{|l|}{{RAPU-R}} & 
\multicolumn{1}{c|}{$0.32$} &
\multicolumn{1}{c|}{$0.38$} &
\multicolumn{1}{c|}{$0.79$} & 
\multicolumn{1}{c|}{$0.86$} 
\\
\cline{3-7}
 \multicolumn{1}{|l|}{} &
 &
 \multicolumn{1}{|l|}{H-CARS} &
\multicolumn{1}{c|}{$\bf{0.37}$} & 
\multicolumn{1}{c|}{$\bf{0.42}$} &
\multicolumn{1}{c|}{$\bf{0.82}$} & 
\multicolumn{1}{c|}{$\bf{0.91}$} 
\\
\cline{3-7}
 \multicolumn{1}{|l|}{} &
 &
 \multicolumn{1}{|l|}{Bandwagon} &
\multicolumn{1}{c|}{$0.09$} & 
\multicolumn{1}{c|}{$0.11$} &
\multicolumn{1}{c|}{$0.16$} & 
\multicolumn{1}{c|}{$0.26$}

\\
\cline{2-7}
\multicolumn{1}{|l|}{} & 
\multicolumn{1}{l}{\multirow{4}{*}{RCF}} & 
\multicolumn{1}{|l|}{{DL-Attack}} & 
 \multicolumn{1}{c|}{$\bf{0.25}$} &
\multicolumn{1}{c|}{$\bf{0.29}$} &
 \multicolumn{1}{c|}{$0.63$} & 
\multicolumn{1}{c|}{$0.71$} 

\\
\cline{3-7}
\multicolumn{1}{|l|}{} &
&
\multicolumn{1}{|l|}{{RAPU-R}} & 
 \multicolumn{1}{c|}{$0.23$} & 
\multicolumn{1}{c|}{$0.28$} &
\multicolumn{1}{c|}{$0.66$} & 
\multicolumn{1}{c|}{$0.73$} 

\\
\cline{3-7}
\multicolumn{1}{|l|}{} &
&
\multicolumn{1}{|l|}{H-CARS} &
 \multicolumn{1}{c|}{$0.24$} & 
\multicolumn{1}{c|}{$\bf{0.29}$} & 
\multicolumn{1}{c|}{$\bf{0.68}$} & 
\multicolumn{1}{c|}{$\bf{0.76}$} 
\\

\cline{3-7}
 \multicolumn{1}{|l|}{} &
 &
 \multicolumn{1}{|l|}{Bandwagon} &
\multicolumn{1}{c|}{$0.06$} & 
\multicolumn{1}{c|}{$0.10$} &
\multicolumn{1}{c|}{$0.13$} & 
\multicolumn{1}{c|}{$0.14$}
\\

\hline
\multicolumn{1}{|l|}{\multirow{8}{*}{\em Yelp}} & 
\multicolumn{1}{l}{\multirow{4}{*}{NCF}} & 
\multicolumn{1}{|l|}{{DL-Attack}} & 
\multicolumn{1}{c|}{$\bf{0.27}$} & 
\multicolumn{1}{c|}{$0.30$} &
 \multicolumn{1}{c|}{$0.63$} & 
\multicolumn{1}{c|}{$0.73$} 
\\
\cline{3-7}
\multicolumn{1}{|l|}{} &
&
\multicolumn{1}{|l|}{{RAPU-R}} & 
\multicolumn{1}{c|}{$0.25$} &
\multicolumn{1}{c|}{$0.31$} &
\multicolumn{1}{c|}{$0.65$} & 
\multicolumn{1}{c|}{$0.74$} 
\\
\cline{3-7}
 \multicolumn{1}{|l|}{} &
 &
 \multicolumn{1}{|l|}{H-CARS} &
\multicolumn{1}{c|}{$0.26$} & 
\multicolumn{1}{c|}{$\bf{0.33}$} &
\multicolumn{1}{c|}{$\bf{0.66}$} & 
\multicolumn{1}{c|}{$\bf{0.76}$} 
\\
\cline{3-7}
 \multicolumn{1}{|l|}{} &
 &
 \multicolumn{1}{|l|}{Bandwagon} &
\multicolumn{1}{c|}{$0.05$} & 
\multicolumn{1}{c|}{$0.08$} &
\multicolumn{1}{c|}{$0.16$} & 
\multicolumn{1}{c|}{$0.21$}

\\
\cline{2-7}
\multicolumn{1}{|l|}{} & 
\multicolumn{1}{l}{\multirow{4}{*}{RCF}} & 
\multicolumn{1}{|l|}{{DL-Attack}} & 
 \multicolumn{1}{c|}{$\bf{0.24}$} &
\multicolumn{1}{c|}{$0.25$} &
 \multicolumn{1}{c|}{$0.56$} & 
\multicolumn{1}{c|}{$0.63$} 

\\
\cline{3-7}
\multicolumn{1}{|l|}{} &
&
\multicolumn{1}{|l|}{{RAPU-R}} & 
 \multicolumn{1}{c|}{$0.22$} & 
\multicolumn{1}{c|}{$0.24$} &
\multicolumn{1}{c|}{$0.59$} & 
\multicolumn{1}{c|}{$0.63$} 

\\
\cline{3-7}
 \multicolumn{1}{|l|}{} &
 &
 \multicolumn{1}{|l|}{H-CARS} &
\multicolumn{1}{c|}{$0.23$} & 
\multicolumn{1}{c|}{$\bf{0.26}$} &
\multicolumn{1}{c|}{$\bf{0.61}$} & 
\multicolumn{1}{c|}{$\bf{0.66}$} 
\\
\cline{3-7}
 \multicolumn{1}{|l|}{} &
 &
 \multicolumn{1}{|l|}{Bandwagon} &
\multicolumn{1}{c|}{$0.03$} & 
\multicolumn{1}{c|}{$0.04$} &
\multicolumn{1}{c|}{$0.12$} & 
\multicolumn{1}{c|}{$0.18$}
\\
\cline{2-7}
\hline
\end{tabular}
\quad
\begin{tabular}{lll|cccc|}
\cline{4-7}
& & & \multicolumn{4}{c|}{{\bf Percentage of Cotrolled Users}}
\\
\hline
 \multicolumn{1}{|l|}{\bf Dataset} & \multicolumn{1}{l}{\thead{Recommendation\\Model}} & \multicolumn{1}{|l|}{\thead{Attack\\Method}} & \multicolumn{1}{c|}{\thead{$0.5\%$}} & \multicolumn{1}{c|}{\thead{$1\%$}} & \multicolumn{1}{c|}{\thead{$3\%$}} & \multicolumn{1}{c|}{\thead{$5\%$}}
\\
\hline
\multicolumn{1}{|l|}{\multirow{8}{*}{\em MovieLens}} & 
\multicolumn{1}{l}{\multirow{4}{*}{NCF}} & 
\multicolumn{1}{|l|}{{DL-Attack}} & 
\multicolumn{1}{c|}{$0.12$} & 
\multicolumn{1}{c|}{$0.14$} &
 \multicolumn{1}{c|}{$0.21$} & 
\multicolumn{1}{c|}{$0.23$} 
\\
\cline{3-7}
\multicolumn{1}{|l|}{} &
&
\multicolumn{1}{|l|}{{RAPU-R}} & 
\multicolumn{1}{c|}{$0.10$} &
\multicolumn{1}{c|}{$0.14$} &
\multicolumn{1}{c|}{$0.23$} & 
\multicolumn{1}{c|}{$0.29$} 
\\
\cline{3-7}
 \multicolumn{1}{|l|}{} &
 &
 \multicolumn{1}{|l|}{H-CARS} &
\multicolumn{1}{c|}{$\bf{0.13}$} & 
\multicolumn{1}{c|}{$\bf{0.18}$} &
\multicolumn{1}{c|}{$\bf{0.29}$} & 
\multicolumn{1}{c|}{$\bf{0.35}$} 
\\
\cline{3-7}
 \multicolumn{1}{|l|}{} &
 &
 \multicolumn{1}{|l|}{Bandwagon} &
\multicolumn{1}{c|}{$0.02$} & 
\multicolumn{1}{c|}{$0.03$} &
\multicolumn{1}{c|}{$0.05$} & 
\multicolumn{1}{c|}{$0.10$}

\\
\cline{2-7}
\multicolumn{1}{|l|}{} & 
\multicolumn{1}{l}{\multirow{4}{*}{RCF}} & 
\multicolumn{1}{|l|}{{DL-Attack}} & 
 \multicolumn{1}{c|}{$\bf{0.06}$} &
\multicolumn{1}{c|}{$0.09$} &
 \multicolumn{1}{c|}{$0.17$} & 
\multicolumn{1}{c|}{$0.22$} 

\\
\cline{3-7}
\multicolumn{1}{|l|}{} &
&
\multicolumn{1}{|l|}{{RAPU-R}} & 
 \multicolumn{1}{c|}{$0.04$} & 
\multicolumn{1}{c|}{$0.08$} &
\multicolumn{1}{c|}{$0.17$} & 
\multicolumn{1}{c|}{$0.21$} 

\\
\cline{3-7}
\multicolumn{1}{|l|}{} &
&
\multicolumn{1}{|l|}{H-CARS} &
 \multicolumn{1}{c|}{$\bf{0.06}$} & 
\multicolumn{1}{c|}{$\bf{0.11}$} & 
\multicolumn{1}{c|}{$\bf{0.20}$} & 
\multicolumn{1}{c|}{$\bf{0.26}$} 
\\

\cline{3-7}
 \multicolumn{1}{|l|}{} &
 &
 \multicolumn{1}{|l|}{Bandwagon} &
\multicolumn{1}{c|}{$0.03$} & 
\multicolumn{1}{c|}{$0.03$} &
\multicolumn{1}{c|}{$0.05$} & 
\multicolumn{1}{c|}{$0.08$}
\\

\hline

\multicolumn{1}{|l|}{\multirow{8}{*}{\em Yelp}} & 
\multicolumn{1}{l}{\multirow{4}{*}{NCF}} & 
\multicolumn{1}{|l|}{{DL-Attack}} & 
 \multicolumn{1}{c|}{$0.05$} &
\multicolumn{1}{c|}{$0.06$} &
 \multicolumn{1}{c|}{$0.09$} & 
\multicolumn{1}{c|}{$0.11$} 

\\
\cline{3-7}
\multicolumn{1}{|l|}{} &
&
\multicolumn{1}{|l|}{{RAPU-R}} & 
 \multicolumn{1}{c|}{$0.05$} & 
\multicolumn{1}{c|}{$0.05$} &
\multicolumn{1}{c|}{$0.09$} & 
\multicolumn{1}{c|}{$0.12$} 

\\
\cline{3-7}
\multicolumn{1}{|l|}{} &
&
\multicolumn{1}{|l|}{H-CARS} &
 \multicolumn{1}{c|}{$\bf{0.06}$} & 
\multicolumn{1}{c|}{$\bf{0.08}$} & 
\multicolumn{1}{c|}{$\bf{0.12}$} & 
\multicolumn{1}{c|}{$\bf{0.16}$} 
\\

\cline{3-7}
 \multicolumn{1}{|l|}{} &
 &
 \multicolumn{1}{|l|}{Bandwagon} &
\multicolumn{1}{c|}{$0.01$} & 
\multicolumn{1}{c|}{$0.02$} &
\multicolumn{1}{c|}{$0.06$} & 
\multicolumn{1}{c|}{$0.08$}
\\
\cline{2-7}

\multicolumn{1}{|l|}{} & 
\multicolumn{1}{l}{\multirow{4}{*}{RCF}} & 
\multicolumn{1}{|l|}{{DL-Attack}} & 
 \multicolumn{1}{c|}{$0.06$} &
\multicolumn{1}{c|}{$0.08$} &
 \multicolumn{1}{c|}{$0.13$} & 
\multicolumn{1}{c|}{$0.15$} 

\\
\cline{3-7}
\multicolumn{1}{|l|}{} &
&
\multicolumn{1}{|l|}{{RAPU-R}} & 
 \multicolumn{1}{c|}{$0.06$} & 
\multicolumn{1}{c|}{$0.09$} &
\multicolumn{1}{c|}{$0.13$} & 
\multicolumn{1}{c|}{$0.16$} 

\\
\cline{3-7}
\multicolumn{1}{|l|}{} &
&
\multicolumn{1}{|l|}{H-CARS} &
 \multicolumn{1}{c|}{$\bf{0.07}$} & 
\multicolumn{1}{c|}{$\bf{0.10}$} & 
\multicolumn{1}{c|}{$\bf{0.15}$} & 
\multicolumn{1}{c|}{$\bf{0.20}$} 
\\

\cline{3-7}
 \multicolumn{1}{|l|}{} &
 &
 \multicolumn{1}{|l|}{Bandwagon} &
\multicolumn{1}{c|}{$0.02$} & 
\multicolumn{1}{c|}{$0.03$} &
\multicolumn{1}{c|}{$0.05$} & 
\multicolumn{1}{c|}{$0.10$}
\\
\cline{2-7}

\hline

\end{tabular}
}
\end{table*}
\subsection{Experimental Setup}
\noindent{\bf Datasets.}
In this paper, we conduct experiments following~\cite{chen2022grease} and \cite{tran2021counterfactual} over two datasets: \textit{Yelp} and \textit{MovieLens100K}.\footnote{Hereinafter, we refer to it simply as \textit{MovieLens}.} \textit{Yelp} contains binary ratings from $31,668$ users on $38,048$ items, while \textit{MovieLens} contains $943$ users' rating scores (in the range $[1,5]$) on $1,682$ movies. According to~\cite{tran2021counterfactual}, we binarize ratings in \textit{MovieLens} as follows: any score greater than $4$ is mapped to $1$, and $0$ otherwise.

\noindent{\bf Target Recommendation Models.}
We experiment with the following target recommendation models:
\begin{itemize}
    \item {\bf \em Neural Collaborative Filtering (NCF)}~\cite{he2017neural}
     replaces the user-item inner product with a neural architecture to capture the complex structure of the user interaction data.
     \item {\bf \em Relational Collaborative Filtering (RCF)}~\cite{xin2019relational}
     is developed to exploit multiple-item relations in recommender systems via a  two-level hierarchical attention mechanism.
\end{itemize}

\noindent{\bf Counterfactual Explanation Method.}
We consider the following CF generation method:
\begin{itemize}
    \item {\bf \em ACCENT}~\cite{tran2021counterfactual} generates CFs for neural recommenders by using influence functions to find items most relevant to a recommendation.
\end{itemize}

\noindent{\bf Baseline Attack Methods.}
We compare our H-CARS attack strategy against the following baselines:
\begin{itemize}
    \item {\bf \em DL-Attack}~\cite{huang2021data} formulates the attack as an optimization problem such that the injected data would maximize the number of normal users to whom the target items are recommended.
    \item {\bf \em RAPU-R}~\cite{zhang2021data} starts from the attack goal and reverses the optimization process to obtain the crafted interactions.
    \item {\bf \em Bandwagon Attack}~\cite{zhang2021data} randomly selects popular items for crafted users who also interact with the target item.
\end{itemize}

\subsection{Attack Setting}


Adapting~\cite{tang2020revisiting}, we sample $5$ target items as $\mathcal{T}$ and limit controlled user selections to $100$ and $15$ for \textit{MovieLens} and \textit{Yelp} respectively, with a learning rate of 0.001. Model extraction parameters are set: $\lambda_1 = 0.76$, $\lambda_2 = 0.0001$ for \textit{MovieLens}; and $\lambda_1 = 0.68$, $\lambda_2 = 0.00001$ for \textit{Yelp}. In the extraction stage, $60\%$ interactions generate CFs per~\cite{tran2021counterfactual}, also used as negative samples for DL-Attack, RAPU-R, and Bandwagon attack. Attackers' performance is assessed using Hit Ratio at $10$ ($HR@10$), measuring the proportion of users with at least one $\mathcal{T}$ item in their top-$10$ recommendations~\cite{li2020sampling}.

\subsection{Analysis of Results}



Our H-CARS attack consistently outperforms or performs similarly to other methods in all target recommendation model and dataset combinations, as shown in Table~\ref{tab:Result80}. For instance, using the \textit{MovieLens} dataset, our method improves over RAPU-R and DL-Attack by $5\%$ and $8\%$, respectively. Notably, our method performs well with limited training data, outperforming RAPU-R by $0.6\%$ with only $30\%$ of the \textit{MovieLens} training data. We observe that our method achieves good performance in the scenario with a limited number of training data. In particular, with only $30\%$ \textit{MovieLens} training data, our method outperforms RAPU-R by $0.6\%$. Similar conclusions can be drawn for other combinations with limited data.  Moreover, as the number of controlled user increases, the performance gap between RAPU-R get enlarged. For instance, compared with RAPU-R, our H-CARS attack achieves $0.1\%$ and $0.6\%$ performance increases on the H-CARS dataset when the percentage of controlled users rises from $0.5\%$ to $5\%$. Overall, our method achieves the best performance compared to other baselines.

\subsection{Ablation Study}
\begin{figure}[ht]
\centering
\includegraphics[width=\columnwidth]{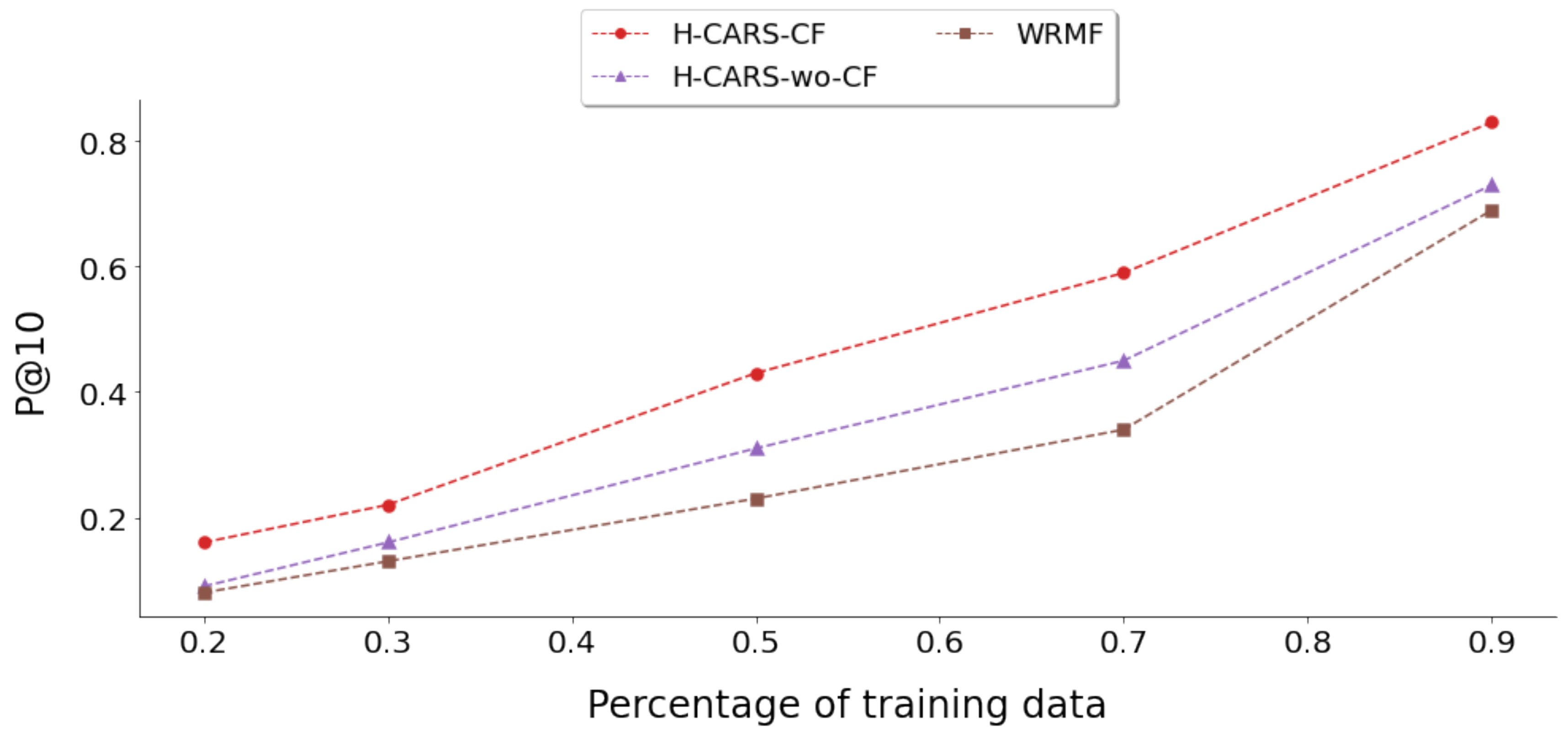}
\caption{The impact of CFs on surrogate models.}
\label{fig:ablation}
 \vspace{-0.5cm} 
\end{figure}

In our study, we evaluate the performance of our model extraction method using the MovieLens dataset and NCF as the original recommender. We apply the $P@10$ precision metric and compare three surrogate models: H-CARS-CF, H-CARS-wo-CF, and WRMF. Results in Figure~\ref{fig:ablation} indicate that H-CARS-CF surpasses the others, validating the effectiveness of $\mathcal{L}_{cf}$-based CFs. Additionally, the performance gap between H-CARS-CF and H-CARS-wo-CF widens with increased training data, highlighting the synergy of CFs and factual data in data augmentation.

\section{Conclusion and Future Work}
\label{sec:conclusion}
We presented a novel approach, H-CARS, that exploits the vulnerabilities induced by counterfactual explanations to launch a poisoning attack on recommender systems. To the best of our knowledge, this is the first such attack proposed in the literature. Our experiments demonstrate that H-CARS is effective, highlighting the importance of considering the security implications of using explainability methods in recommender systems. Future research should explore the potential impact of such attacks on the integrity of the recommender system, as well as develop stronger defenses to mitigate risks in explainable recommender systems.

\begin{acks}
This work was partially supported by projects FAIR (PE0000013) and SERICS (PE00000014) under the MUR National Recovery and Resilience Plan funded by the European Union - NextGenerationEU, the XJTLU Research Development Fund under RDF-21-01-053, TDF21/22-R23-160, Ningbo 2025 Key Scientific Research Programs, Grant/Award Number: 2019B10128, S10120220021, and National Science Foundation 2127918, 2046457 and 2124155.
\end{acks}











\end{document}